\documentclass[12pt]{emulateapj}
\usepackage{graphicx}

\def\kms{{\rm \thinspace km \thinspace s}^{-1}}

\def\Msun{\hbox{$\rm\thinspace M_{\odot}$}}
\def\pc{{\rm\thinspace pc}}

\def\kyr{{\rm\thinspace kyr}}
\def\Myr{{\rm\thinspace Myr}}

\def\K{{\rm\thinspace K}}

\def\ndens{\thinspace\mathrm{cm}^{-3}}

\def\mnras{MNRAS}
\def\apj{ApJ}
\def\aap{A\&A}
\def\apjl{ApJL}
\def\apjs{ApJS}

\def\pasp{PASP}

\def\pasj{PASJ}

\shorttitle{The formation and evolution of the
  Pipe Nebula - is star formation ever isolated?}
\shortauthors{Matthias Gritschneder and Douglas N.C. Lin}

\begin{document}

\title{The role of $\theta$ Oph in the formation and evolution of the
  Pipe Nebula - is star formation ever isolated?}

\author{Matthias Gritschneder$^{1,2}$ and Douglas N.C. Lin$^{1,2}$}
\affil{$^1$Kavli Institute for Astronomy and Astrophysics, Peking
  University,Yi He Yuan Lu 5, Hai Dian, 100871 Beijing, China;\\
$^2$Astronomy and Astrophysics Department, University of California, Santa Cruz,
  CA 95064, USA}
\texttt{gritschneder@ucolick.org}


\begin{abstract}
We propose that the Pipe Nebula is an HII region shell swept up by the
B2 IV $\beta$ Cephei star $\theta$ Ophiuchi. After reviewing the
morphological evidence by recent observations, we perform a series of
analytical calculations. We use realistic HII region parameters derived
with the radiative transfer code Cloudy from observed stellar
parameters. We are able to show that the current size, mass and
pressure of the region can be explained in this scenario.
We investigate the configuration today and come to the conclusion that the
Pipe Nebula can be best described by a three phase medium in pressure
equilibrium. The pressure support is provided by the ionized gas and
mediated by an atomic component to confine the cores at the observed
current pressure.
In the future, star formation in these cores is likely to be either
triggered by feedback of the most massive, gravitationally bound cores
as soon as they collapse or by the supernova explosion of $\theta$
Ophiuchi itself.
\end{abstract}


\section{Introduction}
The Pipe Nebula, first observed by \citet{Onishi:1999lr} is a nearby
molecular cloud region. Due to its relative 
proximity ($D\approx130\pc$, \citealt{Lombardi:2006fk},
$D\approx145\pc$, \citealt{Alves:2007lr}) it provides an
ideal testbed to observe molecular cloud core formation
\citep{Lada:2008qy}. The total spatial extend of the Pipe Nebular is roughly
$14\pc\times3\pc$. The early measurements by \citet{Onishi:1999lr}
showed the gaseous component emitting $^{12}$CO to have a total mass of 
about $10^4\Msun$ and  the component emitting $^{13}$CO to have about
$3\times 10^3\Msun$. In addition, they identified 14 cores emitting
C$^{18}$O. More recent extinction measurements
\citep{Lombardi:2001uq,Lombardi:2006fk} increased the number of cores
to more than 150
\citep{Alves:2007fk,Muench:2007fk,Rathborne:2009uq,Roman-Zuniga:2010kx}. 
The total mass in the upper part of the observed area 
($11\pc\times18\pc$) as inferred from
extinction measurements is found to be $(11 000 \pm 2600)\Msun$
\citep{Lombardi:2006fk} and thus in very good agreement 
with the previous CO observations\footnote{Denote that
  \citet{Onishi:1999lr} assumed a larger distance for the Pipe Nebula.
The mass estimate of \citet{Lombardi:2006fk} would be $\approx
1.7\times10^4\Msun$ at that distance.}. It is a remarkably quiescent
region and therefore often assumed to be the 
ideal case to study isolated star formation. Only in one tip, termed
B59, observations indicate low rate star formation and probably Jeans
fragmentation \citep{Roman-Zuniga:2009uq}. The observed stars in B59
are about $2.6\Myr$ old \citep{Covey:2010vn}. Recent observations with
the Spitzer Space Telescope confirm the extremely low level of star formation in
the Pipe Nebula \citep{Forbrich:2009ys}. Additional observations in the near infrared
\citep{Roman-Zuniga:2009uq,Roman-Zuniga:2010kx}, in NH3 \citep{Rathborne:2008ve} and X-ray
\citep{Forbrich:2010ly} support these findings.
Magnetic field observations \citep{Alves:2008zr,Frau:2010vn,Franco:2010kx} indicate that,
especially in the Stem of the Pipe Nebula, the magnetic field seems to
be aligned perpendicular to the main filament. Very recent 
observations with Herschel \citep{Peretto:2012qc} show that
in the Stem the sub-filaments are more grid-like, whereas in B59 they
are more centrally condensed. 

The Pipe Nebula is located at the edge of the Sco OB2 association. The
closest massive star is the B2 IV $\beta$ Cephei star $\theta$
Ophiuchi (HD 157056) at a projected distance of
about $3\pc$ from the Pipe Nebula. While its variability was known for a
long time, only recently it has been target of detailed
asteroseismologic studies. Non-rigid rotation was proven from
multiplet observation. Observations and models
\citep[e.g.][]{Handler:2005kx,Briquet:2007vn} show that $\theta$ 
Oph is a triple system with an inner binary. 
Stellar models \citep{Lovekin:2010uq} indicate that the brightest star $\theta$
Oph A can be best fit by stellar models with a luminosity of
$log(L/L^C)= 3.75$ and an effective temperature $T_{\rm eff} = 22 590\K$
at an age $t_{\rm star}=15.6\Myr$.
The second closest massive star is the B0 star $\tau$ Sco at a projected
distance of about $20\pc$.

Various models have been proposed to explain the formation and
especially the observed core mass function in the Pipe Nebula \citep[e.g.][]{Heitsch:2009lr}.
The effect of $\theta$ Oph in triggering star formation has been
previously studied by \citet{Onishi:1999lr}. Here, they show that the
stellar winds can not trigger star formation. However, they did not
investigate the effect of the ionizing radiation in the formation and
the current state of the region.
In this work, we interpret the Pipe Nebula as a swept up HII-region
shell. In \S \ref{equations} we review the underlying
physics. In \S \ref{models} we present analytic models and assess the
current state in  \S \ref{currentstate}. We investigate the future evolution
in \S \ref{future} and conclude in \S \ref{conclusions}.

\section{Basic Equations}
\label{equations}
In general, the evolution of an ionized region can be characterized
by two phases. First, the ionization of the so called Stroemgren
sphere \citep{Stromgren:1939fk}, the volume which can be immediately ionized by the star
before the system can react to the increased temperature and
pressure. This sphere has the radius 
\begin{equation}
\label{R_stroem}
R_{\rm S}=\left(\frac{3J_{\rm Ly}}{4\pi n_0^2\alpha_{\rm B}}\right)^{1/3},
\end{equation}
where $J_{\rm Ly}$ is the total ionizing flux of the source, $n_0$ the
number density in the surrounding and $\alpha_{\rm B}$ the sum of the
recombination coefficients for all levels besides the ground stage.
In the second phase, the heated gas reacts to its increase in
temperature and therefore pressure. An approximately isothermal
shock is driven into the surrounding medium. Under the assumption of a
thin shock, the time evolution of the radius is given as 
\begin{equation}
\label{R_t}
R(t)=R_\mathrm{s}\left(1+\frac{7}{4}\frac{a_\mathrm{s,hot}}{R_\mathrm{s}}(t-t_0)\right)^\frac{4}{7},
\end{equation}
where $a_\mathrm{s,hot}$ is the sound speed of the hot, ionized
gas. The time evolution of the density in the ionized region is then
given by
\begin{equation}
\label{rho_t}
\rho(t)=\rho_0\left(1+\frac{7}{4}\frac{a_\mathrm{s,hot}}{R_\mathrm{s}}(t-t_0)\right)^{-\frac{6}{7}}.
\end{equation}
For the accumulated mass we assume that the shell contains all
material swept up from the HII-region, weighted by the current
covering fraction $f$ of the Pipe Nebula for the maximum size of the bubble:
\begin{equation}
\label{M_t}
M(t)=\frac{4\pi}{3}R(t)(\rho_0-\rho(t)) f,
\end{equation}
where 
\begin{equation}
f=\frac{A_{\rm Pipe}}{4\pi R^2_{\rm max}}.
\end{equation}
Here, $A_{\rm Pipe}$ is the current area of the Pipe Nebular tangential
to the massive star and $R_{\rm max}$ is maximal radius of the
HII-region, i.e. the current distance between the star and the nebula.

A more detailed treatment has to include the fact that the region
around $\theta$ Oph is not spherically
symmetric. \citet{Krumholz:2009lr} derived an analytic approximation
for blister-type HII regions. In this case
\begin{equation}
\label{Rb_t}
R_{\rm b}(t)=R_\mathrm{s}\left(\frac{7t}{\sqrt{6}t_\mathrm{s}}\right)^\frac{4}{7},
\end{equation}
where $t\mathrm{s}=R\mathrm{s}/a_\mathrm{s,hot}$, and
\begin{equation}
\label{rhob_t}
\rho_{\rm b}(t)=\rho_0 \left(\frac{7t}{\sqrt{6}t_\mathrm{s}}\right)^{-\frac{6}{7}}.
\end{equation}
$M_{\rm b}(t)$ can then be calculated according to Eq. \ref{M_t}.

\section{Formation of the Pipe Nebula}
\label{models}
\begin{figure}
\begin{center}
\includegraphics[width=10cm]{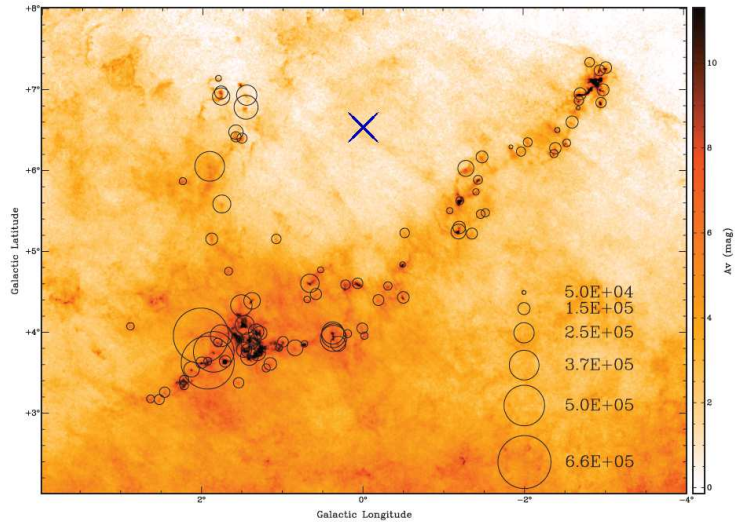}
\caption{Observation of the Pipe Nebula by \citet{Lombardi:2006fk},
  Figure 8 of \citet{Lada:2008qy}. The location of $\theta$
Oph is indicated by the asterisk. The size of the circles indicates
the pressure of the cores (in $\K\ndens$).\label{NICER}}
\end{center}
\end{figure}
Using these basic equations we can now derive a model for the Pipe
Nebula. In the NICER extinction maps (Fig. \ref{NICER},
\citealt{Lombardi:2006fk,Lada:2008qy}), a suggestive geometry can already be
seen. The region around $\theta$ Oph (denoted by the asterisk) provides
significantly less extinction, i.e. is less dense. We interpret this
as the HII region.

In the following, we assume the cold gas to be at $T_{\rm cold}=10\K$
with a mean molecular weight of $\mu_{\rm cold}=1.37$
\citep{Lombardi:2006fk},
corresponding to a sound speed of $a_{\rm
  s,cold}=0.25\kms$. \footnote{Arguably, the gas could be
    initially molecular, e.g. $\mu_{\rm cold}=2.35$. As this just
    corresponds to a different choice of initial $n_0$, we assume it to
    be atomic to enable an easier comparison with the photo-dissociated
  post-shock region later on.}
The temperature in the HII-region is resulting from the balance of heating
by photoelectrons and cooling by forbidden metal lines
\citep[e.g.][]{Osterbrock:1989lr}.

We test three models A, B and C, corresponding to initial number
densities $n_0$ in the cold surrounding of $1\times10^3\ndens$,
$5\times10^3\ndens$ and $1\times10^4\ndens$, respectively. In order to
get a precise estimate for the radius and temperature of the
initial Stroemgren sphere, we employ the radiative transfer code
Cloudy. Calculations were performed with version 08.00 of Cloudy, last
described by \citet{Ferland:1998fk}. We only consider the most
luminous star, $\theta$ Oph A, as it is at least an order
of magnitude brighter than its companions
\citep{Handler:2005kx}\footnote{We also neglect the influence of the
  B0 star $\tau$ Sco, since its projected distance to the Pipe Nebula
  is $\approx20\pc$}. We parametrize $\theta$ Oph A as a black body
with a temperature of $T_{\rm eff} = 
22 590\K$ and a luminosity of ${\rm log}(L/L_\odot)= 3.75$
\citep{Lovekin:2010uq}\footnote{The simulations are performed under
  the assumption of a constant density, solar metallicity and a
  magnetic field of $B=10^{-5}$G}.
The code Cloudy yields $R_{\rm S}=0.094\pc$, $T_{\rm hot}
\simeq 6000\K$ in case A, $R_{\rm S}=0.029\pc$, $T_{\rm hot} \simeq
7000\K$ in case B and $R_{\rm S}=0.017\pc$, $T_{\rm hot} \simeq
7000\K$ in case C. The corresponding sound speeds are calculated 
assuming a molecular weight of $\mu_{\rm hot}=0.55$.

\begin{figure*}
\begin{center}
\includegraphics[width=\textwidth]{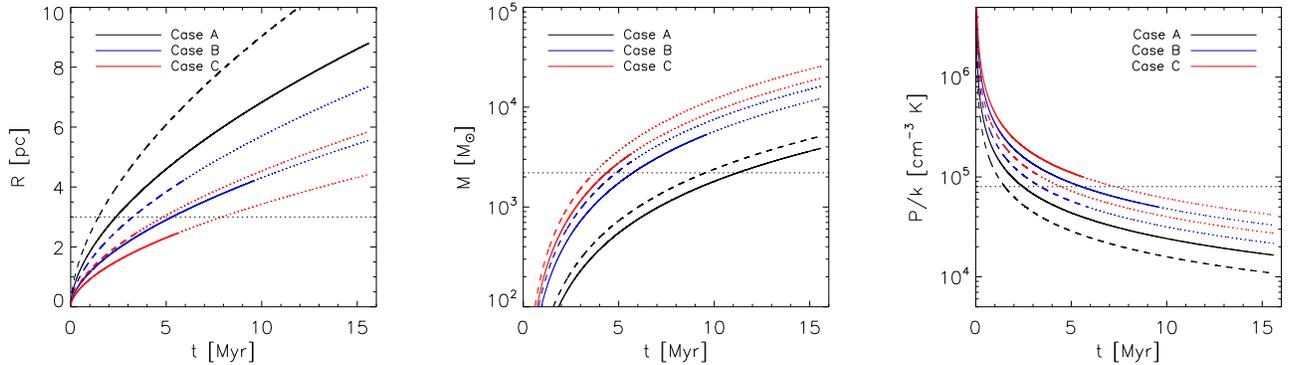}
\caption{Time evolution of the HII-region for the three different
  cases. Solid lines: classical (spherical) HII-region, dashed:
  blister-type HII-region. The lines are continued dotted once the
  shock has reached equilibrium with the ambient surrounding. Dotted
  horizontal lines: current day observational values. Left panel:
  radius of the shell. Center: swept up mass in a Pipe Nebula sized
  region. Right: pressure in the hot, ionized gas. \label{HII_evol}} 
\end{center}
\end{figure*}

In Fig. \ref{HII_evol} we show the evolution in all three cases according
to Eqn. \ref{R_t}, \ref{rho_t}, \ref{Rb_t} and \ref{rhob_t}.
Solid lines correspond to the classical (spherical) HII region, dashed
lines correspond to the blister case. The dotted horizontal line is
the value inferred from observations. It can be directly seen that in all
models the final radius is bigger than $3\pc$. This is especially
true for the blister case. However, recent numerical simulations by
\citet{Gendelev:2012ss} show that in the blister case the analytical
approximation overestimates the radius.  In their simulations it is only
bigger by about $15\%$ than the spherical estimate, while the analytic
prediction would be $20\%$. 
In both cases the value is of order of the
observed value and at the distance of the Pipe Nebula
($\approx130-146\pc$) the discrepancy can be a projection effect, as
Eqn. \ref{R_t} and \ref{Rb_t} give the real radius, while we can only
observe the projected radius.

Next, we take a look at the mass in the swept-up shell.  We
estimate the surface of the Pipe Nebula tangential to the direction
of the star to be the same as the projected size of the Nebula,
i.e. $A_{\rm Pipe}=14\pc\times3\pc$. This is of course ad hoc and only
an order of magnitude estimate. However, panel 2 of
Fig. \ref{HII_evol} shows clearly that in the HII-region scenario it
is no problem to accumulate enough mass to be consistent with the
observed value in the Pipe Nebula ($M\approx2.2\times 10^3\Msun$,
  see \S \ref{currentstate}).

Another constraint for the models is the pressure required to
confine the observed cores. \citet{Lada:2008qy} estimate the required
average pressure at the surface of the cores to be $\approx 8\times
10^4\K\ndens$. In our model, this pressure has to be initially
supplied by the ionized gas (for a detailed discussion see \S
\ref{currentstate}). From Fig. \ref{HII_evol} it can be seen that it is
challenging to maintain this pressure in our model for the entire
time-span of $15.6\Myr$. Nevertheless, the analytical value is close to the
observed value in all three cases. In addition, a closer investigation
of Fig. \ref{NICER} shows, that the cores in the swept up shell have
lower pressures, whereas the cores in the central region have
higher pressure. We discuss this further in \S \ref{currentstate}.

The shock front will stagnate as soon as the pressure in the neutral gas equals the
pressure in the ionized gas. This is reached in Cases B and C. Once
the front stagnates, the corresponding lines in Fig. \ref{HII_evol}
are continued dotted. All three quantities are remarkably close to the
observations at this point. This leads to a
  straightforward interpretation. The HII region expanded until
  stagnation. At the same time, part of the shocked gas and the post-shock gas expand,
  as they are preheated by the photodissociating photons of the B-star
to about $100\K$ (see \ref{currentstate}). The increased motion
($a_{\rm s,warm}=0.78\kms$) leads to a pile-up of material towards the
solid boundary of the ionized gas and to a broadening of the
shock in the opposite direction. Thus, the shock broadens with half
$a_{\rm s,warm}$, reaching about $3.2\pc$ in the $8\Myr$ since
stagnation in Case B. The density at stagnation in the cold shock is
$n_0$ and will drop according to the increase in temperature. In case
B, with a new temperature of $100\K$, this corresponds to
$500\ndens$. Taking into account the additional pile up this is very
close to the value inferred from the observations ($n_{\rm atomic} =
774\ndens$, see below).

In addition, the equations in \S \ref{equations} are derived under the
approximation of a thin shock. This approximation will fail
eventually.  Then, the speed of the ionization 
front will be lower than the analytical estimate. Therefore, the
radius and swept up mass are best viewed as an upper
limit. Correspondingly, the pressure is a lower limit, since a smaller
region will lead to higher densities and thus to a higher pressure in
the ionized gas.

\section{Current State}
\label{currentstate}
We now continue to discuss the current state of the Pipe Nebula in more
detail. The most obvious constraints are the mass, pressure and
current size.

To asses the current mass, it is necessary to take a detailed look at
the observations. The total mass in the upper part of the observed
region ($b>+3^{\circ}$) is inferred to be $10^4\Msun$ from extinction
measurements by \citet{Lombardi:2006fk}. They provide a detailed
comparison with the previous $^{12}{\rm CO}$-measurements by \citet{Onishi:1999lr}
and find encouraging agreement. To estimate the mass inside the Pipe
Nebula itself it is first necessary to look at the mass in the dense
component alone. From \citet{Lombardi:2006fk} Fig. 7, it is reasonable
to assume that the dense component has an extinction $A_K>0.4$. Inside
this isoextinction contour there is about $40\%$ of the total
mass \citep[][their Fig. 27]{Lombardi:2006fk}. About half of this mass is
in the Pipe Nebula, therefore the denser
component of the nebula itself has a mass of about $2.2\times10^3\Msun$. This is
in agreement with the $3\times10^3\Msun$ in the entire region inferred
by \citet{Onishi:1999lr} from $^{13}{\rm CO}$, which traces the denser
gas (e.g. their Fig. 5). The total mass in the cores is about
$250\Msun$ \citep[][their Table 1]{Lada:2008qy}.

From these masses, we can derive the structure of the Nebula. The
cores have a known pressure of $\approx 8\times10^{4}\K\ndens$. The ionized gas
can supply this pressure. Independent of the uncertainties in \S
\ref{models}, we perform a Cloudy simulation on the current
static state. It shows, that $\theta$ Oph A can ionize gas at
$n_0=8\ndens$ up to a radius of $2.75\pc$ to $T=5000\K$. With
$\mu_{\rm ion}=0.55$ the pressure is then $7.3\times10^{4}\K\ndens$,
which is a bit less than the observed average density of the
cores. However, a closer look at Fig. \ref{NICER} reveals, that the
cores along the rim generally have a lower pressure (i.e. smaller
circles), whereas the high pressure cores are towards the central region. A straightforward
assumption is that the atomic component in between the cores and the
ionized gas is in pressure equilibrium with both, acting as a
buffer. Assuming a volume of  $14\pc\times3\pc\times2\pc$ for the Pipe
Nebula, and $\mu=1.37$ \citep{Lombardi:2006fk}, this corresponds to
$n_{\rm atomic} = 774\ndens$. To be in pressure equilibrium, the atomic
phase has to be at $T=100K$. Thus, the bulk of the mass
($\approx87.5\%$) is in the atomic phase. To keep the medium at this
temperature, significant heating is required. The heating can be supplied
by the diffuse radiation. In addition, \citet{Arthur:2011fk} showed
that for B-stars the photo-dissociation-region (PDR) is wider than
for O-stars. This is due to the peak at lower energies in the
spectrum. Therefore, in B-stars, the ratio of photo-dissociating
photons to ionizing photons is higher. As a consequence, the
photo-dissociation, i.e. the disruption of ${\rm H}_2$ molecules, outruns
the ionization front, leading to a warmer region at about $100\K$ with
a thickness of about 30\% of the radius of the HII region.

Concerning the high pressure cores in the central region, it is probably best to
assume that this region is currently undergoing gravitational
collapse. Possibly, the swept up HII-shell encountered a denser region
and the resulting density enhancement lead to a gravitational
unstable region. Thus, the cores in the region are gravitationally
bound.

The width of the Pipe Nebula can be explained by assuming the
  shock and post-shock region expand after stagnation with $\frac{1}{2}a_{\rm
    s,warm}$ (see \S \ref{models}). Therefore, the Case C would lead
  to $4.4\pc$ and Case B to $3.2\pc$, as observed.

\section{Further Evolution and Implications for the IMF}
\label{future}
As the bulk of the mass is in the atomic phase, this phase will most
likely determine the future evolution of the Pipe Nebula. A possible
scenario is triggered star formation from within. When
the most massive clumps turn into stars, they will ionize and thereby
heat the atomic phase. This will in turn increase the pressure on
the cold cores, forcing them to form stars.

In the context of the transition from the Core Mass Function (CMF) to
the Initial Mass Function (IMF) it is very interesting to look at the
increase of the ambient pressure quantitatively. If the radiation of
the newborn stars increases the background pressure by a factor of
two, the shift from the IMF to the CMF could be explained
\citep{Lada:2008qy}. 

There are three cores of about $10\Msun$ and one with $\approx
20\Msun$ in the Pipe Nebula. Assuming that the entire most massive
cores end up in a single stars, the further evolution will be similar
to our model A. As can be directly seen from Fig. \ref{HII_evol},
there would be a shock proceeding through the Pipe
nebula. However, due to the number density in the atomic gas, the
evolution of this shock would take several $\Myr$, whereas star
formation in general seems to be more synchronized. For the $10\Msun$
cores, the evolution would be even slower. Furthermore, this
triggering scenario would leave the top end of the IMF unaffected as
these cores turn into stars before they themselves increase the
ambient pressure.

A more violent scenario will occur if $\theta$ Oph explodes in a
supernova before the massive cores turn into stars. Given the
relatively small distance, the Pipe Nebula will be hit by a blast wave
in the Sedov phase within a few $\kyr$ after the explosion. Mostly
likely, the smaller cores will get disrupted, while the bigger cores
will get triggered into collapse within another few $\kyr$
\citep[see e.g.][]{Gritschneder:2012fc}. This could indeed explain the
small age spread observed in star forming regions.

Another environmental effect is the influence of the Sco OB2
association. As pointed out previously \citep[e.g.][]{de-Geus:1992fk} and discussed recently by
 \citet{Peretto:2012qc}, the star forming region termed B59 is pointing towards that
association and resembles a bow shock like structure. However, the age
estimate for the young stars in B59 of $2.6\Myr$
\citep{Covey:2010vn} suggests that this interaction triggered the stars in
the past and is not currently driving star formation further down the stem.

\section{Discussion \& Conclusions}
\label{conclusions}
There are several uncertainties involved in the model. The main
uncertainty lies in the assumption of a thin shock for the analytic
solution. Once the shock thickens, it will lead to a smaller
region and a less dense shell with a higher pressure, as discussed
above. Another possibility to constrain the HII region to the current
size would be magnetic fields. \citet{Krumholz:2007fk} were able to
demonstrate that the evolution of a shell stagnates earlier for a
magnetized medium since the Alfven speed is higher than the sound
speed \citep[see also][]{Arthur:2011fk}.

Another uncertainty lies in the assumed initial conditions for the
Cloudy models, e.g. the assumed abundances, grains and magnetic fields
will influence the outcome. However, for the temperature and the size
of the HII region, these simulations give more realistic answers than
the simple assumptions leading to Eq. \ref{R_t}. Furthermore, the most
important input parameter - besides the stellar parameters taken
directly from observations - is still the density, as it enters with
$n_0^2$.

The assumed time scale for the sweeping up of the Pipe Nebula is
$\approx15\Myr$. This is comparable to the free-fall time of a cloud so a
remaining question is why the nebula did not collapse. One possible
solution might be the finding of \citet{Arthur:2011fk} that B-stars
have an extended PDR. In this broad region behind the ionization front
the temperature is close to 100K. As e.g. shown by
\citet{Gritschneder:2010uq}, a higher temperature in the cold gas is
prohibitive of structure formation and thus hindering
collapse. 

Besides these uncertainties, the scenario of $\theta$ Oph swiping up
the Pipe Nebula presented in this paper can, especially in Case B,
successfully explain: 
\begin{enumerate}
\item{The observed morphology of the Pipe Nebula, especially the
    spheroidal shape of the less dense region.}
\item{The current width, as a shock broadening with 
    $\frac{1}{2}a_{\rm s,warm} =0.44\kms$ will reach a width of $3.2\pc$ in
    the $8\Myr$ since the stagnation.}
\item{The current mass and size can be easily reached.}
\item{More importantly, the pressure to confine the cores can be supplied.}
\end{enumerate}
Especially the pressure of the cores is otherwise puzzling. Up to
now, the only possible explanation for this confinement was the
self-gravity of the cloud. This is highly unlikely, as the cores in a
self-gravitating system are the first instances to react to the
collapse and therefore should be bound, whereas most of the cores are
observed to be unbound.

\section{Acknowledgements}
M.G. acknowledges funding by the Alexander von Humboldt Foundation in
form of a Feodor-Lynen Fellowship and by the China
National Postdoc Fund Grant No. 20100470108 and the National Science
Foundation of China Grant No. 11003001. D.N.C.L. acknowledges funding
by the NASA grant NNX08AL41G.

\bibliographystyle{apj}

\begin{thebibliography}{32}
\expandafter\ifx\csname natexlab\endcsname\relax\def\natexlab#1{#1}\fi

\bibitem[{{Alves} \& {Franco}(2007)}]{Alves:2007lr}
{Alves}, F.~O., \& {Franco}, G.~A.~P. 2007, \aap, 470, 597

\bibitem[{{Alves} {et~al.}(2008){Alves}, {Franco}, \& {Girart}}]{Alves:2008zr}
{Alves}, F.~O., {Franco}, G.~A.~P., \& {Girart}, J.~M. 2008, \aap, 486, L13

\bibitem[{{Alves} {et~al.}(2007){Alves}, {Lombardi}, \& {Lada}}]{Alves:2007fk}
{Alves}, J., {Lombardi}, M., \& {Lada}, C.~J. 2007, \aap, 462, L17

\bibitem[{{Arthur} {et~al.}(2011){Arthur}, {Henney}, {Mellema}, {de Colle}, \&
  {V{\'a}zquez-Semadeni}}]{Arthur:2011fk}
{Arthur}, S.~J., {Henney}, W.~J., {Mellema}, G., {de Colle}, F., \&
  {V{\'a}zquez-Semadeni}, E. 2011, \mnras, 414, 1747

\bibitem[{{Briquet} {et~al.}(2007){Briquet}, {Morel}, {Thoul}, {Scuflaire},
  {Miglio}, {Montalb{\'a}n}, {Dupret}, \& {Aerts}}]{Briquet:2007vn}
{Briquet}, M., {Morel}, T., {Thoul}, A., {Scuflaire}, R., {Miglio}, A.,
  {Montalb{\'a}n}, J., {Dupret}, M.-A., \& {Aerts}, C. 2007, \mnras, 381, 1482

\bibitem[{{Covey} {et~al.}(2010){Covey}, {Lada}, {Rom{\'a}n-Z{\'u}{\~n}iga},
  {Muench}, {Forbrich}, \& {Ascenso}}]{Covey:2010vn}
{Covey}, K.~R., {Lada}, C.~J., {Rom{\'a}n-Z{\'u}{\~n}iga}, C., {Muench}, A.~A.,
  {Forbrich}, J., \& {Ascenso}, J. 2010, \apj, 722, 971

\bibitem[{{de Geus}(1992)}]{de-Geus:1992fk}
{de Geus}, E.~J. 1992, \aap, 262, 258

\bibitem[{{Ferland} {et~al.}(1998){Ferland}, {Korista}, {Verner}, {Ferguson},
  {Kingdon}, \& {Verner}}]{Ferland:1998fk}
{Ferland}, G.~J., {Korista}, K.~T., {Verner}, D.~A., {Ferguson}, J.~W.,
  {Kingdon}, J.~B., \& {Verner}, E.~M. 1998, \pasp, 110, 761

\bibitem[{{Forbrich} {et~al.}(2009){Forbrich}, {Lada}, {Muench}, {Alves}, \&
  {Lombardi}}]{Forbrich:2009ys}
{Forbrich}, J., {Lada}, C.~J., {Muench}, A.~A., {Alves}, J., \& {Lombardi}, M.
  2009, \apj, 704, 292

\bibitem[{{Forbrich} {et~al.}(2010){Forbrich}, {Posselt}, {Covey}, \&
  {Lada}}]{Forbrich:2010ly}
{Forbrich}, J., {Posselt}, B., {Covey}, K.~R., \& {Lada}, C.~J. 2010, \apj,
  719, 691

\bibitem[{{Franco} {et~al.}(2010){Franco}, {Alves}, \&
  {Girart}}]{Franco:2010kx}
{Franco}, G.~A.~P., {Alves}, F.~O., \& {Girart}, J.~M. 2010, \apj, 723, 146

\bibitem[{{Frau} {et~al.}(2010){Frau}, {Girart}, {Beltr{\'a}n}, {Morata},
  {Masqu{\'e}}, {Busquet}, {Alves}, {S{\'a}nchez-Monge}, {Estalella}, \&
  {Franco}}]{Frau:2010vn}
{Frau}, P. {et~al.} 2010, \apj, 723, 1665

\bibitem[{{Gendelev} \& {Krumholz}(2012)}]{Gendelev:2012ss}
{Gendelev}, L., \& {Krumholz}, M.~R. 2012, \apj, 745, 158

\bibitem[{{Gritschneder} {et~al.}(2010){Gritschneder}, {Burkert}, {Naab}, \&
  {Walch}}]{Gritschneder:2010uq}
{Gritschneder}, M., {Burkert}, A., {Naab}, T., \& {Walch}, S. 2010, \apj, 723,
  971

\bibitem[{{Gritschneder} {et~al.}(2012){Gritschneder}, {Lin}, {Murray}, {Yin},
  \& {Gong}}]{Gritschneder:2012fc}
{Gritschneder}, M., {Lin}, D.~N.~C., {Murray}, S.~D., {Yin}, Q.-Z., \& {Gong},
  M.-N. 2012, \apj, 745, 22

\bibitem[{{Handler} {et~al.}(2005){Handler}, {Shobbrook}, \&
  {Mokgwetsi}}]{Handler:2005kx}
{Handler}, G., {Shobbrook}, R.~R., \& {Mokgwetsi}, T. 2005, \mnras, 362, 612

\bibitem[{{Heitsch} {et~al.}(2009){Heitsch}, {Ballesteros-Paredes}, \&
  {Hartmann}}]{Heitsch:2009lr}
{Heitsch}, F., {Ballesteros-Paredes}, J., \& {Hartmann}, L. 2009, \apj, 704,
  1735

\bibitem[{{Krumholz} \& {Matzner}(2009)}]{Krumholz:2009lr}
{Krumholz}, M.~R., \& {Matzner}, C.~D. 2009, \apj, 703, 1352

\bibitem[{{Krumholz} {et~al.}(2007){Krumholz}, {Stone}, \&
  {Gardiner}}]{Krumholz:2007fk}
{Krumholz}, M.~R., {Stone}, J.~M., \& {Gardiner}, T.~A. 2007, \apj, 671, 518

\bibitem[{{Lada} {et~al.}(2008){Lada}, {Muench}, {Rathborne}, {Alves}, \&
  {Lombardi}}]{Lada:2008qy}
{Lada}, C.~J., {Muench}, A.~A., {Rathborne}, J., {Alves}, J.~F., \& {Lombardi},
  M. 2008, \apj, 672, 410

\bibitem[{{Lombardi} \& {Alves}(2001)}]{Lombardi:2001uq}
{Lombardi}, M., \& {Alves}, J. 2001, \aap, 377, 1023

\bibitem[{{Lombardi} {et~al.}(2006){Lombardi}, {Alves}, \&
  {Lada}}]{Lombardi:2006fk}
{Lombardi}, M., {Alves}, J., \& {Lada}, C.~J. 2006, \aap, 454, 781

\bibitem[{{Lovekin} \& {Goupil}(2010)}]{Lovekin:2010uq}
{Lovekin}, C.~C., \& {Goupil}, M.-J. 2010, \aap, 515, A58

\bibitem[{{Muench} {et~al.}(2007){Muench}, {Lada}, {Rathborne}, {Alves}, \&
  {Lombardi}}]{Muench:2007fk}
{Muench}, A.~A., {Lada}, C.~J., {Rathborne}, J.~M., {Alves}, J.~F., \&
  {Lombardi}, M. 2007, \apj, 671, 1820

\bibitem[{{Onishi} {et~al.}(1999){Onishi}, {Kawamura}, {Abe}, {Yamaguchi},
  {Saito}, {Moriguchi}, {Mizuno}, {Ogawa}, \& {Fukui}}]{Onishi:1999lr}
{Onishi}, T. {et~al.} 1999, \pasj, 51, 871

\bibitem[{{Osterbrock}(1989)}]{Osterbrock:1989lr}
{Osterbrock}, D.~E. 1989, {Astrophysics of gaseous nebulae and active galactic
  nuclei} (University Science Books)

\bibitem[{{Peretto} {et~al.}(2012){Peretto}, {Andr{\'e}}, {K{\"o}nyves},
  {Schneider}, {Arzoumanian}, {Palmeirim}, {Didelon}, {Attard}, {Bernard}, {Di
  Francesco}, {Elia}, {Hennemann}, {Hill}, {Kirk}, {Men'shchikov}, {Motte},
  {Nguyen Luong}, {Roussel}, {Sousbie}, {Testi}, {Ward-Thompson}, {White}, \&
  {Zavagno}}]{Peretto:2012qc}
{Peretto}, N. {et~al.} 2012, \aap, 541, A63

\bibitem[{{Rathborne} {et~al.}(2009){Rathborne}, {Lada}, {Muench}, {Alves},
  {Kainulainen}, \& {Lombardi}}]{Rathborne:2009uq}
{Rathborne}, J.~M., {Lada}, C.~J., {Muench}, A.~A., {Alves}, J.~F.,
  {Kainulainen}, J., \& {Lombardi}, M. 2009, \apj, 699, 742

\bibitem[{{Rathborne} {et~al.}(2008){Rathborne}, {Lada}, {Muench}, {Alves}, \&
  {Lombardi}}]{Rathborne:2008ve}
{Rathborne}, J.~M., {Lada}, C.~J., {Muench}, A.~A., {Alves}, J.~F., \&
  {Lombardi}, M. 2008, \apjs, 174, 396

\bibitem[{{Rom{\'a}n-Z{\'u}{\~n}iga} {et~al.}(2010){Rom{\'a}n-Z{\'u}{\~n}iga},
  {Alves}, {Lada}, \& {Lombardi}}]{Roman-Zuniga:2010kx}
{Rom{\'a}n-Z{\'u}{\~n}iga}, C.~G., {Alves}, J.~F., {Lada}, C.~J., \&
  {Lombardi}, M. 2010, \apj, 725, 2232

\bibitem[{{Rom{\'a}n-Z{\'u}{\~n}iga} {et~al.}(2009){Rom{\'a}n-Z{\'u}{\~n}iga},
  {Lada}, \& {Alves}}]{Roman-Zuniga:2009uq}
{Rom{\'a}n-Z{\'u}{\~n}iga}, C.~G., {Lada}, C.~J., \& {Alves}, J.~F. 2009, \apj,
  704, 183

\bibitem[{{Str{\"o}mgren}(1939)}]{Stromgren:1939fk}
{Str{\"o}mgren}, B. 1939, \apj, 89, 526

\end{thebibliography}

\end{document}